\documentclass[12pt]{article}
\usepackage{latexsym}
\newcommand{\Poin}{Poincar\'e}
\def\Re{{\rm Re ~}}
\def\Im{{\rm Im ~}}
\def\diag{{\rm diag ~}}

 \newcommand{\ft}[2]{{\textstyle\frac{#1}{#2}}}
\newsavebox{\zzzbar}
\sbox{\zzzbar}
  {\setlength{\unitlength}{0.9em}
  \begin{picture}(0.6,0.7)
  \thinlines
  \put(0,0){\line(1,0){0.6}}
  \put(0,0.75){\line(1,0){0.575}}
  \multiput(0,0)(0.0125,0.025){30}{\rule{0.3pt}{0.3pt}}
  \multiput(0.2,0)(0.0125,0.025){30}{\rule{0.3pt}{0.3pt}}
  \put(0,0.75){\line(0,-1){0.15}}
  \put(0.015,0.75){\line(0,-1){0.1}}
  \put(0.03,0.75){\line(0,-1){0.075}}
  \put(0.045,0.75){\line(0,-1){0.05}}
  \put(0.05,0.75){\line(0,-1){0.025}}
  \put(0.6,0){\line(0,1){0.15}}
  \put(0.585,0){\line(0,1){0.1}}
  \put(0.57,0){\line(0,1){0.075}}
  \put(0.555,0){\line(0,1){0.05}}
  \put(0.55,0){\line(0,1){0.025}}
  \end{picture}}

\newsavebox{\uuunit}
\sbox{\uuunit}
    {\setlength{\unitlength}{0.825em}
     \begin{picture}(0.6,0.7)
        \thinlines
        \put(0,0){\line(1,0){0.5}}
        \put(0.15,0){\line(0,1){0.7}}
        \put(0.35,0){\line(0,1){0.8}}
       \multiput(0.3,0.8)(-0.04,-0.02){12}{\rule{0.5pt}{0.5pt}}
     \end {picture}}

\newcommand{\rmi}{{\rm i}}

\csname @addtoreset\endcsname{equation}{section}

plus 0.2pt minus 0.1pt \textheight = 45\baselineskip
\advance\textheight by \topskip \textwidth 465pt  \columnsep 10pt
 \newcommand{\Ka}{K\"ahler}
 \newcommand{\Nphi}{{\cal N} }
 
\newcommand{\Yrho}{Y}

\begin{document}
\begin{titlepage}
\begin{flushright}
KUL-TF-99/12\\ SU-ITP-99/19\\ CITA-99-16\\ hep-th/9907124
\end{flushright}
\vspace{.5cm}
\begin{center}
\baselineskip=16pt {\LARGE \bf    Gravitino Production After
Inflation
}\\
\vfill

\

{\large Renata Kallosh,$^1$ Lev Kofman,$^2$ Andrei Linde,$^1$
and Antoine Van Proeyen$^{3,\dagger}$ 
  } \\
\vfill

\

{\small $^1$ Department of Physics, Stanford University, Stanford,
CA 94305, USA
\\ \vspace{6pt}
$^2$ CITA, University of Toronto, 60 St George Str, Toronto, ON
M5S 3H8, Canada
\\ \vspace{6pt}
${}^3$ Instituut voor Theoretische Fysica, Katholieke
 Universiteit Leuven,\\
Celestijnenlaan 200D B-3001 Leuven, Belgium }
\end{center}
\vfill
\begin{center}
{\bf Abstract}
\end{center}
{\small We investigate the production of gravitinos in a
cosmological background. Gravitinos can be produced during
preheating after inflation due to a combined effect of
interactions with an oscillating inflaton field and absence of
conformal invariance. In order to get insight on conformal
properties of gravitino we reformulate phenomenological
supergravity in $SU(2,2|1$)-symmetric way. The Planck mass and F-
and D-terms appear via the gauge-fixed value of a superfield that
we call conformon. We find that in general the probability of
gravitino production is {\it not}\, suppressed by the small
gravitational coupling. This may lead to a copious production of
gravitinos after inflation. Efficiency of the new non-thermal
mechanism of gravitino production is very sensitive to the choice
of the underlying theory. This may put strong constraints on
certain classes of inflationary models. }
\vspace{2mm} \vfill
\hrule width 3.cm {\footnotesize \noindent $^\dagger$
Onderzoeksdirecteur, FWO, Belgium }
\end{titlepage}

\section{Introduction}


The possibility of excessive production of gravitinos  is one of
the most complicated problems of cosmological models based on
supergravity. Such particles decay very late, and lead to
disastrous cosmological consequences unless the ratio of the
number density of gravitinos $n_{3/2}$ to the entropy density $s$
is extremely small. For example, the ratio of the number density
of gravitinos $n_{3/2}$ to the entropy density $s$ should be
smaller than $O(10^{-14})$ for gravitinos  with mass $O(100)$ GeV
\cite{gravitinos,Moroi}. The standard thermal mechanism of
gravitino production involves scattering of particles at high
temperature in the early universe.  To avoid excessive production
of gravitinos one must assume that the reheating temperature of
the universe after inflation was smaller than  $10^8 - 10^9$ GeV
\cite{gravitinos,Moroi}.

However, gravitinos can be produced not only in the thermal bath
after reheating, but  even earlier, during  the oscillations of
the inflaton field at the end of inflation. We already know that
bosons as well as spin 1/2 fermions can be copiously produced by
the coherently oscillating inflaton field. Quite often this effect
occurs in a non-perturbative way during the stage of preheating
\cite{KLS, GK}. Similar effect may occur for gravitinos. According
to \cite{FKL}, nonthermal gravitino/moduli production may rule out
certain classes of inflationary models which otherwise would be
quite legitimate.

 The theory of the cosmological gravitino production is very
complicated. Recently production of transversal gravitino
components (helicity 3/2) was studied in Ref. \cite{maz} (see also
\cite{moduli} where an attempt has been made to study this
question using perturbation theory). In this paper we will
investigate production of all gravitino components, transversal
(helicity 3/2) and longitudinal (helicity 1/2). As we will show,
 the production rate of the longitudinal
gravitino component can be much greater than that of the
transversal gravitino components.\footnote{In flat spacetime,
transversal components, which exist even if supersymmetry is
unbroken, correspond to the helicity 3/2, while longitudinal
components with non-vanishing $\psi_0$ and $\gamma^i\psi_i$
correspond to the helicity 1/2. Although the helicity concept has
a less precise meaning in FRW metrics, we still will use this
loose definition as a shortcut to the transversal and longitudinal
components.}

Since gravitino is a part of the gravitational multiplet, one
could expect that their production must be strongly suppressed by
the small gravitational coupling. Indeed, usually the production
of particles occurs because their effective mass changes
nonadiabatically
during the oscillations of the inflaton field \cite{KLS}. This is
the main effect responsible for the production of  the gravitinos
with helicity 3/2. The gravitino mass $m$ at small values of the
inflaton field $\phi$ is proportional to $M_P^{-2}W$, where $W$ is
a superpotential. Thus the amplitude of the oscillations of the
gravitino mass is suppressed by $M_P^{-2}$. That is why production
of  the gravitino components with helicity 3/2 is relatively
inefficient \cite{maz,moduli}.

There is another mechanism to be considered, which is related to
breaking of conformal invariance \cite{FKL}.   It is well known,
for example, that expansion of the universe does not lead to
production of massless vector particles and massless fermions of
spin 1/2 because the theory of such particles is conformally
invariant and the Friedmann universe is conformally flat.
Meanwhile, massless scalar particles minimally coupled to gravity
(as well as gravitons) are created in an expanding universe
because the theory describing these particles is not conformally
invariant. The rate of scalar particle production is determined by
the Hubble constant $H \equiv {{\dot a} \over a}= M_P^{-1}
 \sqrt{{\rho/ 3}}$, where $\rho$ is the energy density. If
 similar effects are possible for gravitinos, they could be
much stronger than the effects discussed in the previous
paragraph. Indeed,  $H$ is suppressed only by the first degree of
$M_P^{-1}$. As a result $H$ typically is much greater than
$m_{3/2}$ after inflation, so its time-dependence may lead to a
more efficient particle production.

The issue of conformal invariance of gravitinos is rather
nontrivial, and until now it has not been thoroughly examined. For
gravitinos with helicity 3/2 the effects proportional to $H$ do
not appear, and therefore the violation of conformal invariance
for such particles is very small, being proportional to their mass
\cite{maz}. However, as we will show, the theory of gravitinos
with helicity 1/2 is not conformally invariant, and therefore such
particles will be produced during the expansion of the universe
even if one neglects their mass.

But the most surprising effect which we have found is that the
production of the gravitinos of helicity 1/2 by the oscillating
scalar field $\phi$ in general is {\it not} suppressed by any
powers of $M_P^{-1}$, and therefore their production can be very
efficient. The magnitude of the effect is model-dependent. For
example,  this effect does not occur in the simplest model of a
single chiral multiplet with a quadratic effective potential
$m_{\phi}^2\phi^2/2$. Meanwhile this effect is very strong in the
theory with the effective potential $\lambda \phi^4/4$. Gravitinos
in this theory are produced very quickly, within about ten
oscillations of the inflaton field, with occupation numbers
$n_k\sim 1/2$ over a large range of momenta $k \leq
\sqrt{\lambda}\phi$,
 which is very different from usual perturbative production.

This result may have important cosmological implications, since it
may allow to rule out certain classes of cosmological theories.
The nature of this effect resembles the well known fact that the
longitudinal components of massive vector bosons at high energy
behave in the same way as the Goldstone boson  that was eaten by
the vector field \cite{Corn}. A similar effect is known to exist
in the theory of technicolor \cite{lenny}. A more direct analogy
is the transmutation of gravitational interactions of gravitinos
with helicity 1/2 to weak interactions found by Fayet
\cite{Fayet}. In our case the effect is nonperturbative, and its
adequate interpretation is achieved by finding solutions of the
gravitino equations in a nontrivial self-consistent cosmological
background.

In order to study conformal properties of gravitinos we
reformulated the standard N=1 phenomenological supergravity in an
SU$(2,2|1)$-invariant way, which makes the conformal properties of
the theory manifest and explains how the conformal symmetry is
broken. Our formulation describes arbitrary number of chiral and
vector multiplets and is flexible enough to allow investigation of
regimes where the superpotential $W$ vanishes.

In application to the theory of gravitino production, we
concentrate on the simplest models with one chiral multiplet
$\Phi$ and  arbitrary superpotential. We present classical
equations of motion and constraints for the transverse and
longitudinal gravitino in the expanding Friedmann universe
interacting with the moving inflaton field. We use the gauge
where the goldstino is absent. Then we solve classical equations
for gravitino. This solution confirms the generic prediction from
the $SU(2,2|1)$-symmetric theory that the  longitudinal gravitinos
are not conformal.

We represent equations describing gravitino components with
helicities 3/2 and 1/2 in a form analogous to the equations for
the usual spin 1/2 fermions with time-dependent mass. This allows
to reduce, to a certain extent, the problem of gravitino
production to the problem of production of particles with spin 1/2
after preheating \cite{GK}.

Finally, we estimate the number density of gravitinos produced by
the oscillating scalar field in several inflationary models, and
show that in some models the ratio $n_{3/2}/s$ may substantially
exceed the  bound $n_{3/2}/s \leq O(10^{-14})$.

A detailed account of our investigation will be given in a
separate publication \cite{long}. Here we will only outline the
main points of our study and present the most interesting results.

\section{Supergravity Lagrangian and conformal properties of
gravitino}

Fundamental M-theory, which should encompass both supergravity and
string theory, at present experiences rapid changes. One may still
expect that the low-energy physics will be described by the N=1
d=4 supergravity
 \cite{supergravity}
and  address the issues of the early universe cosmology
 in the context of the most general phenomenological N=1
supergravity--Yang--Mills-matter theory \cite{general}.

We are interested in conformal properties of supergravity fields,
which include various spin fields,  in the conformally flat FRW
metric describing the early universe:
\begin{equation}
 g_{\mu\nu}(x)= a^2 (\eta) \eta_{\mu\nu}\,.
\end{equation}
In particular, we will be interested in conformal properties of
gravitino. Supergravity is not a conformally invariant theory,
despite
 the fact that it has a long history of being derived using
the superconformal tensor calculus \cite{SU221} as a technical
tool. Therefore it is difficult even to address this issue as the
supergravity fields do not have specific conformal weights. To
solve this problem, the idea is to view  the supergravity theory
as a  gauge-fixed version of the conformally invariant theory
describing the  most
 general  $N=1$ gauge theory superconformally
coupled to supergravity \cite{long}.  The derivation $SU(2,2|1)$
invariant Lagrangian of $N=1$ supergravity coupled to $n+1$ chiral
multiplets (with complex scalars $X_I$ and fermions $\Omega _I$)
and Yang--Mills vector multiplets (with gauginos $\lambda ^\alpha
$ and vectors $W_\mu ^\alpha $) superconformally will be presented
in \cite{long}. It has no dimensional parameters. It consists of 3
parts, depending, respectively, on a real function ${\cal N}$, a
holomorphic function $\mathcal{W}$ and a gauge group 2-tensor
$f_{\alpha \beta }$. Each of them is conformally invariant by
itself.
\begin{equation}
{\cal L}_{superconf} =  [\Nphi (X, \bar X)]_D + [\mathcal{W}(X)]_F
+ \left[f_{\alpha \beta }(X) \bar \lambda_L ^\alpha \lambda_L
^\beta\right] _F \label{symbL} \ .
\end{equation}
The statement of the $SU(2,2|1)$ symmetry of the action
(\ref{symbL}) includes, among others, the symmetries under the
following set of local dilatations, with parameter $\sigma (x)$,
 for the metric and gravitino, for
the scalars and spinors of the chiral multiplets, and for the
vectors and spinors of the gauge multiplet, respectively:
\begin{eqnarray}\label{weyl}
 g_{\mu\nu}'&=& e^{-2\sigma(x) } g_{\mu\nu}\ , \qquad
 \psi'_\mu =
e^{-\frac{1}{2}\left[\sigma (x)\right]}\psi _\mu\,,
\nonumber\\\label{chiral}
 X_I' &=&
e^{\sigma (x)} X_I\ , \qquad  \qquad \Omega_I' =
e^{\frac{3}{2}\sigma (x)} \Omega_I\ , \nonumber\\
 W_\mu ^{\alpha '}&=& W_\mu ^\alpha \ , \qquad  \qquad
  \qquad \lambda^{\alpha'}= e^{\frac{3}{2}\sigma (x)}
  \lambda^{\alpha}\,.
\label{gauge}
\end{eqnarray}
The  function of scalars ${\cal N} (X, \bar X)$ codifies the
information on \Ka\ manifold. The holomorphic function of scalars
${\cal W}(X)$ codifies the superpotential. They transform as
follows under local dilatations:
\begin{equation}
{\cal N}(X', \bar X') = e^{2\sigma(x) }{\cal N} (X, \bar
X)\,,\qquad {\cal W}(X') = e^{3\sigma(x) }{\cal W} (X)\,,\qquad
f_{\alpha \beta }(X')=f_{\alpha \beta }(X)\,.
\end{equation}
The important term in the conformal action which allows us to
distinguish between conformal properties of helicity $\pm 3/2$ and
$\pm 1/2$ gravitino is the following:
\begin{eqnarray}
  [\Nphi ]_De^{-1}={1\over 6}\Nphi (X, \bar X) \left ( R + \bar
\psi_\mu R^\mu+ e^{-1}
\partial _\mu (e \bar \psi \cdot \gamma \psi ^\mu) \right)  +
\dots\,.
\end{eqnarray}
The gauge fixing of the local dilatation of the conformally
invariant action presented in \cite{long} leads to the
standard\footnote{ In fact two different gauges can be used to fix
the $R$-part of the superconformal symmetry.  With the first
choice ${\cal W} ={\cal W}^*$ we get the N=1 phenomenological
supergravity \cite{general} depending on ${\cal G}$ (with
$ e^{-{\cal G}}\equiv M_P^{-6} e^{\cal K}|W|^2$), while the
second one $\Yrho= \Yrho^*$ (see notation below) gives the version which is non-singular
in the limit of  the vanishing superpotential $\mathcal{W}=0$.
This version is closer to the one in \cite{BW} which was obtained
by the superspace methods. The limit $\mathcal{W}=0$ from the
first version has been discussed at the end of \cite{Rel}.
However, we find the second version more suitable for cosmology.
It will be presented below.} Poincar\'e supergravity theory.

The  dimensionful constants, Plank mass and F- and D-terms, appear
via the gauge fixed value of the conformal compensator superfield,
which we call {\it conformon}. The original
 $n+1$
complex variables $X_I$ are split into one complex scalar
conformon field $\Yrho $,  and $n$ physical complex scalars $z_i$,
which are hermitian coordinates for parametrizing the \Ka\
manifold in the
 \Poin\ theory. One defines
\begin{equation}
  X_I=\Yrho\, x_I(z_i)\ ,
\label{Xrhoxz}
\end{equation}
and the local dilatation takes the form in which only the
conformon $\Yrho$ transforms and the physical scalars do not
transform under the local dilatation
\begin{equation}
\Yrho' = e^{\sigma (x)} \Yrho \ , \qquad z_i' = z_i\,.
\label{conf}\end{equation} In these variables the gauge-fixing of
the dilatational invariance (\ref{gauge}),
 (\ref{conf}) is given by
\begin{equation}
 {\cal N} (\Yrho,\bar \Yrho, z, \bar z)  = -  3  |\Yrho|^2  \exp \left(-
\ft{1}{3}{\cal K}(z, \bar z) \right)= - 3 M_P^2 , \label{rho}
\end{equation}
where the first equation defines ${\cal K}(z, \bar z)$, which is the \Ka\
potential, and the second is the gauge fixing. Here $M_P \equiv
M_{Planck}/\sqrt{8\pi} \sim 2 \times 10^{18}$ GeV. Thus the
conformon field $\Yrho$ is frozen to $|\Yrho| = M_P  \exp
\left(\ft{1}{6}{\cal K}(z, \bar z)\right) $, i.e. it becomes a functional
of the  \Ka\ potential and not an independent field. The theory
becomes that of the Poincar\'e supergravity theory with
\begin{eqnarray}
  [\Nphi ]_De^{-1}=- {1\over 2} M_P^2 \left ( R + \bar
\psi_\mu R^\mu
 \right)  +\dots\,.
\end{eqnarray}
Here $  R^\mu=\gamma ^{\mu \rho \sigma }{\cal D} _\rho \psi
_\sigma
 $,
where ${\cal D} _\rho$ is a covariant derivative. The  local
dilatation of the metric and gravitino in (\ref{weyl}) is not
compensated anymore by the local dilatation of scalars,
\begin{equation}
 \Yrho = \Yrho' \neq  e^{\sigma(x) }\Yrho\ , \qquad
{\cal N} (X, \bar X)= {\cal N}(X', \bar X') \neq
 e^{2\sigma(x) }{\cal N} (X, \bar X)\,.
\end{equation}
The same happens with the ${\cal W}$ part of the theory:
\begin{equation}
{\cal W} (X)= \Yrho^3M_P^{-3} W(z)
  \qquad \Longrightarrow \qquad  {\cal W} (X)= {\cal W} (X') \neq
  e^{3\sigma(x) }{\cal W}   (X)\,.
\end{equation}
We have chosen here to give $W$ mass dimension~3.

Thus, after the freezing of the conformon field some part of the
transformation cannot be performed and therefore some parts of the
phenomenological supergravity Lagrangian are not invariant under
dilatations. One can try to change the dilatational weight for
these fields to compensate the appearance of powers of $M_P$.
However, this does not help, since the terms with derivatives on
the conformon field are absent after the gauge fixing. The
gravitino field equation which follows from the superconformal
action is
\begin{equation}
 R^\mu -  \gamma^\mu  \psi ^\nu \partial _\nu \ln \sqrt {- {\cal N}}
+  \gamma  \cdot   \psi \partial _\mu \ln \sqrt {- {\cal N}} +
\dots =0\,.
\end{equation}
In the FRW cosmological problems only time derivatives of the
 scalar fields are
important, therefore in $\psi ^\nu\partial _\nu \ln \sqrt {- {\cal
N}}$ only the term $\tilde \psi ^0\partial _0 \ln\sqrt {- {\cal
N}}$ is relevant. After gauge fixing the conformal symmetry will
be broken for configurations for which either
\begin{equation}
 \gamma  \cdot  \psi \neq 0 \qquad \rm or  \qquad  \psi_0\neq 0\,.
\end{equation}
Only such terms will be sensitive to the absence of the terms
$\partial _0 \ln\sqrt {- {\cal N}}$ due to gauge
fixing\footnote{In \cite{GP}, where an attempt to study conformal
properties of gravitino has been made, it was assumed that
 $\gamma \cdot   \psi=\psi^0 =0$ and conformal symmetry of gravitino
was
deduced in the context of pure supergravity.  Without scalars,
however, pure supergravity  does not support a cosmological
background. In the presence of matter the assumption that
 $\gamma \cdot   \psi=\psi^0 =0$ is not valid
and  conformal symmetry is broken.} when $-{\cal N}=3 M_P^2$. The
gravitino in the general theory with spontaneously broken
supersymmetry will be massive. The states of a free massive spin
3/2 particle were studied by Auvil and Brehm in \cite{AB} (see
also \cite{Moroi} for the nice review). A free massive gravitino
has  $\gamma  \cdot \psi =0$. Helicity $\pm 3/2$ states are given
by transverse space components of gravitino, $\psi_i^T$. Helicity
$\pm 1/2$ states are given by the time component of the gravitino
field $\psi_0$. In cases when gravitino interacts with gravity and
other fields, we will find that
 $\gamma
\cdot  \psi \neq 0$. It will be a function of $\psi_0$. Thus the
consideration of superconformal symmetry lead us to a conclusion
that helicity $\pm 1/2$ states of gravitino are not conformally
coupled to the metric. When these states are absent, the $\pm 3/2$
helicity states are conformally coupled (up to the mass terms, as
usual). Thus the conformal properties of gravitino are simple,  as
it is known for scalars: if the action has  an additional term
${1\over 12} \phi^2 R$, the massless scalars are conformal. If
this term is absent, the scalars are not conformal. Note that both
these statements are derivable from the superconformal action. We
will see the confirmation of this prediction in the solutions of
the gravitino equations below.

As we already explained, our formulation starting with
superconformal action \cite{long} provides  flexibility in the
choice of the form of $N=1$ phenomenological supergravity. If we
take ${\cal W}= {\cal W}^*$ gauge for $R$-symmetry, we get the
action \cite{general} depending on the combination of the
K\"{a}hler potential and the superpotential, called ${\cal G}$.
Here we use the $\Yrho= \Yrho^* $ gauge for $R$-symmetry and present
the form of the phenomenological Lagrangian in which the
K\"{a}hler potential and the superpotential are not combined in
one function. This allows to avoid problems which sometimes appear
when the superpotential $W$ vanishes. The action can be written as
\begin{eqnarray}
e^{-1}{\cal L}&=&-\ft12M_P^2\left[ R + \bar \psi_\mu R^\mu+{\cal
L}_{SG,torsion}\right]
-M_P^2g_i{}^j\left[(\hat{\partial }_\mu z^i)(\hat{\partial }^\mu
z_j)+
 \bar \chi _j  \not\!\! {\cal D} \chi^i+ \bar \chi^i
 \not\!\! {\cal D} \chi_j \right]\nonumber\\
&+&(\Re f_{\alpha \beta})\left[ -\ft14 F_{\mu \nu }^\alpha F^{\mu
\nu \,\beta } -\ft12 \bar \lambda ^\alpha \not\!\!\hat{\cal
D}\lambda ^\beta \right] +\ft 14\rmi(\Im f_{\alpha \beta})\left[
 F_{\mu \nu }^\alpha \tilde F^{\mu \nu \,\beta }- \hat{\partial
}_\mu\left( \bar \lambda ^\alpha \gamma _5\gamma ^\mu \lambda
^\beta\right)\right] \nonumber\\
 &-&M_P^{-2} e^{\cal K}\left[ -3 WW^*+({\cal D}^iW) g^{-1}
 {}_i{}^j({\cal D}_jW)\right] -\ft{1}{2}(\Re f)^{-1\,\alpha \beta }
 {\cal P}_\alpha {\cal P}_\beta
 \nonumber\\
&+&\ft18(\Re f_{\alpha \beta})\bar \psi _\mu \gamma ^{\nu \rho
}\left( F_{\nu \rho }^\alpha+ \hat F_{\nu \rho }^\alpha \right)
\gamma ^\mu \lambda ^\beta \nonumber\\ &+& \left\{M_P^2g _j{}^i
\bar \psi_{\mu L}(\hat{\not\! \partial } z^j)
 \gamma^\mu \chi_i
- \ft14f^i_{\alpha \beta}\bar \chi _i\gamma ^{\mu \nu } \hat
F_{\mu \nu }^{-\alpha } \lambda _L^\beta \right. \nonumber\\
 &&\ + \ft12 e^{{\cal K}/2} W\bar \psi _{\mu R} \gamma ^{\mu \nu }\psi
_{\nu R}
 + \bar  \psi_R  \cdot \gamma\left[\ft12\rmi\lambda_L^\alpha
 {\cal P}_\alpha
+\chi _i  e^{{\cal K}/2} {\cal D}^iW  \right] \nonumber\\
 && \
 - e^{{\cal K}/2} ({\cal D}^i{\cal D}^j  W)\bar \chi _i\chi _j
 +\ft 12\rmi(\Re f)^{-1\,\alpha \beta} {\cal P}_\alpha f^i_{\beta
\gamma }
\bar \chi _i\lambda ^\gamma
  - 2 M_P^2\xi _\alpha{}^i g_i{}^j  \bar \lambda ^\alpha \chi_j
 \nonumber\\
  &&\ + \ft14 M_P^{-2} e^{{\cal K}/2}({\cal D}^j W) g^{-1}{}_j{}^i
  f_{\alpha \beta i}\bar \lambda_R ^\alpha \lambda_R ^\beta
 \nonumber\\
&&\left.\ - \ft14f^i_{\alpha \beta}\bar \psi _R\cdot\gamma \chi _i
\bar \lambda _L^\alpha \lambda _L^\beta
 +\ft14 ({\cal D}^i\partial ^jf_{\alpha \beta })\bar \chi _i\chi_j
  \bar \lambda _L^\alpha
 \lambda _L^\beta + h.c.\right\}\label{phenomL}\\
 &+&M_P^2g_j{}^i\left( \ft{1}{8}e^{-1}\varepsilon
^{\mu\nu\rho\sigma}\bar \psi _\mu \gamma
 _\nu \psi _\rho \bar \chi ^j\gamma _\sigma \chi _i
 -\bar \psi _\mu \chi ^j\,\bar \psi^\mu \chi _i\right)
\nonumber\\ &+&M_P^2\left( R_{ij}^{k\ell }-\ft12
g_i{}^kg_j{}^\ell\right) \bar \chi^i\chi ^j\bar \chi _k\chi
_\ell\nonumber\\ & +&\ft 3{64}M_P^{-2}\left( (\Re f_{\alpha \beta
}) \bar \lambda ^\alpha
  \gamma _\mu \gamma _5\lambda ^\beta\right)^2
- \ft1{16}M_P^{-2}f^i_{\alpha \beta }\bar \lambda_L ^\alpha
\lambda_L ^\beta g^{-1}{}_i{}^jf_{\gamma \delta j} \bar \lambda_R
^\gamma \lambda_R ^\delta \nonumber\\ &+&\ft18(\Re f)^{-1\,\alpha
\beta} \left( f^i_{\alpha \gamma }\bar \chi _i\lambda ^\gamma-
f_{\alpha \gamma i }\bar \chi ^i\lambda ^\gamma \right) \left(
f^j_{\beta \delta}\bar \chi _j\lambda ^\delta -f_{ \beta \delta
j}\bar \chi ^j\lambda ^\delta \right) \nonumber \,.
\end{eqnarray}
The $L$ and $R$ denote left and right chirality, e.g.\ $\lambda
_L=\ft12(1+\gamma _5)\lambda $, while for the $\chi $, the
chirality is indicated by the position of the index: $\chi _i$ is
left chiral, while $\chi ^i$ is right chiral. For the scalars,
$z^i$ is the complex conjugate of $z_i$. The \Ka\ metric is
$g_i{}^j$, which is used also for covariant derivatives and \Ka\
curvature
\begin{equation}
  \Gamma_i^{jk} = g^{-1}{} _i^\ell \partial ^jg^k_\ell\,,\qquad
   R_{ij}^{k\ell }\equiv g_i^m\partial _j \Gamma _m^{k\ell}\,.
\label{Kaconnection}
\end{equation}
Extra $i$ indices on quantities, e.g.\ $f_{\alpha \beta }^i$
denote derivatives, here the derivative of $f_{\alpha \beta }$
with respect to $z_i$. Other covariant derivatives and notations
are (antisymmetrization $[\mu \nu ]$ with weight~1, metric
signature $(-,+,+,+)$, notation $\partial ^i=\frac{\partial
}{\partial z_i}$
 and $\gamma _5=\rmi\gamma _0\gamma _1\gamma _2\gamma _3$)
\begin{eqnarray}
\hat{F}^\alpha_{\mu \nu }&=&F^\alpha_{\mu \nu }+\bar \psi _{[\mu
}\gamma _{\nu ]}\lambda ^\alpha\,,\qquad F^\alpha_{\mu \nu
}=2\partial _{[\mu }W^\alpha _{\nu ]}+W_\mu ^\beta W_\nu ^\gamma
f_{\beta \gamma }^\alpha \ , \nonumber\\ \tilde F^{\mu \nu\,\alpha
}&=& \ft12 e^{-1}\varepsilon ^{\mu \nu \rho \sigma }
F^\alpha_{\rho \sigma }\,,\qquad \varepsilon^{0123}=\rmi\,,\qquad
\hat{\partial }_\mu  =  \partial _\mu -W_\mu ^\alpha \delta
_{\alpha }\ , \nonumber\\ R^\mu &=&\gamma ^{\mu \rho \sigma }{\cal
D}_\rho \psi _\sigma \,,\qquad {\cal D}_{[\mu }\psi _{\nu
]}=\left( \partial _{[\mu} +\ft14 \omega _{[\mu} {}^{ab}(e)\gamma
_{ab} +\ft \rmi 2 A^B_{[\mu} \gamma _5\right)\psi _{\nu ]}\ ,
\nonumber\\ \hat{\cal D}_\mu \lambda ^\alpha &=&\left[ \partial
_\mu +\ft14 \left( \omega _\mu {}^{ab}(e)+\ft14\left(2 \bar \psi
_\mu \gamma ^{[a}\psi ^{b]}+\bar \psi ^a\gamma _\mu \psi
^b\right)\right) \gamma _{ab} +\ft \rmi 2 A^B_\mu \gamma _5\right]
\lambda ^\alpha -W_\mu ^\gamma \lambda ^\beta f_{\beta \gamma
}^\alpha \ , \nonumber\\ {\cal D}_\mu \chi _i&=&\left( \partial
_\mu +\ft14\omega _\mu
  {}^{ab}(e)\gamma _{ab}-\ft \rmi 2 A_\mu ^B\right) \chi _i+\Gamma
_i^{jk}\chi
  _j \hat{\partial }_\mu z_k  -W_\mu ^\alpha  \chi _j \partial
  ^j\delta _\alpha z_i \ ,\nonumber\\
 A_\mu ^B&=&\ft \rmi 2 \left[(\partial_i{\cal K})  \hat{\partial
}z^i-(\partial
^i{\cal K})\hat{\partial
  }z_i+3W_\mu ^\alpha ( r^*_\alpha-r_\alpha )\right]\ ,
\label{allconventions}
\\
{\cal D}^iW&=&\partial^iW+ W\partial^i{\cal K}\,,\qquad
 {\cal D}^i{\cal D}^jW\equiv \partial ^i{\cal D}^jW+
  \left( \partial ^i {\cal K}\right) {\cal D}^jW-\Gamma ^{ij}_k{\cal
D}^kW\,,\nonumber
\end{eqnarray}
where $\delta _\alpha $ is  a symbol for the transformation under
the gauge group for all fields. For the conformon field $\Yrho$ and
for the rest of the scalars $z_i$ we have
\begin{equation}
  \delta _\alpha \Yrho =\Yrho\, r_\alpha (z)\,,\qquad
  \delta _\alpha z_i= \xi_{\alpha i}(z)\ ,
\label{delarhoz}
\end{equation}
where $r$ and $\xi $ are $n+1$ holomorphic functions for every
symmetry, such that the quantities in (\ref{symbL}) are invariant.
That determines also
\begin{equation}
  {\cal P}_\alpha(z,z^*)=
\rmi\,M_P^2\left(  \xi_{\alpha i}(z)\partial
^i{\cal K}(z,z^*)-3r_\alpha(z)\right) = \rmi\,M_P^2\left( -\xi _\alpha
{}^i\partial _i{\cal K}(z,z^*)+3r_\alpha^*(z^*)\right) \,. \label{bosD}
\end{equation}
See \cite{long} for details.

The appearance of $M_P= |\Yrho|e^{-{\cal K}/6} $ in various places in this
Lagrangian shows that the conformal symmetry is broken. One can
rescale the fields with $M_P$ so that they have standard kinetic
terms. For our purpose it will be convenient to  replace the
scalar field $ z^i $ by ${\Phi^i \over M_P}$, chiral fermions
$\chi_i$ by ${\chi_i \over M_P}$, and similar for the gravitino,
$\psi_\mu \to {\psi_\mu \over M_P}$.

\section{Gravitino equations}

In general background metrics in the presence of complex scalar
fields with non-vanishing VEV's,  the starting equation for the
gravitino  has in the left hand side  the kinetic part $R^{\mu}$
 and a rather lengthy right hand
side  which will be given in \cite{long}.   Apart of varying
gravitino mass $m=M_P^{-2}e^{{\cal K} \over 2} W$, the right hand side
contains a chiral
 connection $A^B_\mu $ (see (\ref{allconventions})) and various
mixing terms
like those in the 4th, 5th and 6th lines of the phenomenological
Lagrangian (\ref{phenomL}). For a self-consistent setting of the
problem, the gravitino equation should be supplemented by the
equations for the fields mixing with gravitino, as well as by the
equations determining the gravitational background and the
evolution of the scalar fields.


Let us make some simplifications. We consider the supergravity
multiplet and a single chiral multiplet
 containing
a complex scalar field $z={\Phi\over M_P}$ with a superpotential
$W$
 and a single chiral fermion $\chi$.
This is a simple non-trivial extension which allows to study
gravitino in the non-trivial FRW  cosmological metric
 supported by
the scalar field.
 A nice feature of this model is that
   the chiral fermion $\chi$
can be gauged to zero so that the mixing between $\psi_{\mu}$ and
$\chi$ in (\ref{phenomL}) is absent. We also can choose the
non-vanishing VEV of the scalar field in the real direction, $\Re
\Phi = {\phi\over \sqrt 2}$,\,  $\Im \Phi=0$,  so that $A^B_\mu
=0$. The field $\phi = \sqrt 2\ \Re \Phi$ plays the role of the
inflaton field.\footnote{Typical time evolution of the homogeneous
inflaton field
starts with the regime of inflation when $\phi$  slowly
 rolls  down.
One can construct  a superpotential $W$ which provides
 chaotic inflation for $\phi > M_P$.
When  $\phi(t)$ drops below  $ \simeq  M_P$, it begins to oscillate
  coherent oscillations around the minimum
of its the effective potential $V(\phi)$.}
 Then from (\ref{phenomL})  we can obtain  the master
equation for the gravitino field
\begin{equation}
 \not\!\! {\cal D} \psi _\mu + m \psi _\mu  =
  \left( {\cal D}_\mu -\frac{m}{2}\gamma_\mu\right) \gamma^\nu
  \psi_\nu\,,
   \label{master}
\end{equation}
where gravitino mass $m=m(\phi(\eta))$ is given by
\begin{equation}
m = e^{{\cal K}/2}\, {W\over M_P^2} \ .
 \label{grmass}
\end{equation}
Gravitino equation (\ref{master}) is a curved spacetime
generalization of the familiar gravitino equation
 $(\not\!
{\partial} +m_0) \psi_{\mu}=0$
 in a
flat metric, where $m_0$ is a constant gravitino mass.

 The generalization  of the constraint  equations
$\partial^{\mu}\psi_{\mu}=0$ and   $\gamma^{\mu} \psi_{\mu}=0$
 reads as
\begin{eqnarray}
&&{\cal D}^\mu\psi_\mu-\not\!\! {\cal D} \gamma^{\mu} \psi_{\mu} +
\ft32 m \gamma^\mu \psi_\mu =0 \ ,\label{constr3}\\ &&\ft32
m^2\gamma^\mu \psi_\mu+ m'a^{-2}\gamma^0\gamma^i \psi_i=
-\ft{1}{2}G_{\mu \nu }\gamma^\mu \psi^\nu \,,\qquad G_{\mu \nu
}\equiv R_{\mu \nu }-\ft12 g_{\mu \nu }R\,, \label{constr1}
\end{eqnarray}
where $'$ stands for a conformal time derivative
${\partial_{\eta}}$. It is important that the covariant derivative
in these equations must include both the spin connection and the
Christoffel symbols, otherwise equation ${\cal D} _\mu
\gamma_\nu=0$ used for the derivation of these equations is not
valid.

 The last equation will be especially important for us.   Naively,
one could expect that in the limit $M_P\to \infty$, gravitinos
should completely decouple from the background. However, this
equation implies that this is not the case for the gravitinos with
helicity 1/2. Indeed, from (\ref{constr1}) one can find an
algebraic relation between $ \gamma^0\psi_0$ and $\gamma^i
\psi_i$:
\begin{equation}
\gamma ^0\psi_{0}=\hat A  \gamma ^i \psi_i   \ . \label{A}
\end{equation}
Here $\hat A$ is a matrix
 which will  play a
 crucial role in our  description of the interaction of
gravitino with the varying background fields. If $\rho$ and $p$
are the background energy-density and pressure, we have $G^0_0=
M^{-2}_P\rho$,\, $ G^i_k=-  M^{-2}_P p\, \delta^i_k$, and one can
represent the matrix $\hat A$ as follows:
\begin{equation}
  \hat A= {{ p-3m^2M^2_P}
 \over { \rho+ 3  m^2M^2_P }}\,
+\,\gamma_0{{2 m' a^{-1}M^2_P}
 \over { \rho+ 3  m^2 M^2_P}} =A_1+\gamma _0A_2\,.
\label{A2}
\end{equation}

Note that in the limit $M_P \to \infty$ with fixed $\Phi $ in
$z=\Phi M_P^{-1}$, one has $m = M_P^{-2}W$. If $W$ does not blow
up in this limit, this matrix $\hat{A}$ is given by
\begin{equation}
  \hat A= {{ p} \over { \rho}} +\gamma_0{{2 \dot W}
 \over { \rho}}   \ ,
\label{A3}
\end{equation}
where $\dot{}$ stands for derivative  $\partial_t$, and the
relation between physical and conformal times is given by
$dt=a(\eta)d\eta$. In the limit of flat case without moving
scalars, $A=-1$.

For definiteness, we will consider the minimal K\"{a}hler
potential ${\cal K}= z z^*= {\Phi \Phi^*\over M_P^2}$. In the models
where the energy-momentum tensor is determined by the energy of a
classical scalar field and $\Phi$ depends only on time we have
\begin{equation}\label{Gscalar}
\rho = |\dot \Phi|^2 +V\,, ~~~~  p = |\dot \Phi|^2 -V\,, ~~~~
V(\Phi) = e^{\cal K}\, \left(\left|\partial_\Phi W+ {\Phi^*\over {M_P^2}}
W\right|^2-{ 3\over M_P^2} {\left|W\right|^2} \right)\,.
\end{equation}
We will use the representation of gamma matrices where
$\gamma_0=\diag(\rmi,\rmi,-\rmi,-\rmi)$. Then in (\ref{A2}) the
 combination $A=A_1+iA_2$ emerges. For a single
chiral multiplet we obtain $|A|=1$.  (One can show that $|A|=1$ for
the theories with one chiral multiplet   even if the K\"{a}hler is
not minimal.)
Therefore
 $A$  can be represented as\footnote{Initial conditions at
inflation at $\eta \to -\infty$ correspond to $p=-\rho$, $m'=0$
and $A=-1$,
 which gives $\mu(-\infty)=0$. Alternatively, we can start with
inflaton oscillations at $\eta=0$, which defines the phase up to
some constant.
 The final results  depend only on $\mu$.}
\begin{equation}
  A =-\exp \left(2{\rm i}\int_{-\infty }^t dt \,\mu (\eta)\right)
\,.
\label{Astarmu}
\end{equation}
Using the Einstein equations, one obtains for $\mu $ (for minimal
\Ka\ potential, and real scalar field):
\begin{eqnarray}
 \mu &=&M_P^{-2}e^{{\cal K}/2}\left({\cal D}{\cal D}W+W\right)
  +3(HM_P^{-1}\dot \Phi -mm_1)\frac{m_1}{M_P^{-2}\dot \Phi^2
  +m_1^2}~, \nonumber\\
  &&m_1\equiv m'=M_P^{-2} e^{{\cal K}/2}{\cal D}W\,,\qquad
  H\equiv \dot aa^{-1}= a'a^{-2}\,.
\label{muism2}
\end{eqnarray}

The expression for $\mu$ becomes much simpler and its
interpretation is more transparent if the amplitude of
oscillations of the field $\Phi$ is much smaller than $M_P$. In
the limit $\Phi/M_P \to 0$ one has
\begin{eqnarray}\label{gm}
  \mu = \partial_\Phi \partial_\Phi W \ .
\end{eqnarray}
This coincides  with the mass of both fields of the chiral
multiplet (the scalar field and spin 1/2 fermion) in rigid
supersymmetry. When supersymmetry is spontaneously broken, the
chiral fermion, goldstino, is `eaten' by gravitino which becomes
massive and acquires helicity $\pm 1/2$ states in addition to
helicity $\pm 3/2$ states of the massless gravitino.

The matrix $\hat A$ does not become constant in the limit $M_P \to
\infty$. The phase (\ref{Astarmu}) rotates when the background
scalar field oscillates. The amplitude and sign of $A$ change two
times within each oscillation. Consequently, the relation between
$\gamma ^0\psi_{0}$ and $ \gamma ^i \psi_i$ also oscillates during
the field oscillations. This means that the gravitino with
helicity 1/2 (which is related to $\psi_0$) remains coupled to the
changing background even in the limit $M_P \to \infty$. In a
sense, the gravitino with helicity 1/2 remembers its goldstino
nature. This is the main reason why the gravitino production in
this background in general is not suppressed by the gravitational
coupling. The main dynamical quantity which is responsible for the
gravitino production in this scenario will not be the small
changing gravitino mass $m(t)$, but the mass of the chiral
multiplet $\mu$, which is much larger than $m$. As we will see,
this leads to efficient production of gravitinos in the models
where the mass of the `goldstino' nonadiabatically changes with
time.

We shall  solve the
 master equation (\ref{master}) using the constraint equations
in the form (\ref{constr1}) and (\ref{constr3}). We use plane-wave
ansatz $\psi_{\mu} \sim e^{\rmi {\bf k \cdot x}}$ for the
space-dependent part. Then $\psi_i$ can be decomposed\footnote{We
use now $\psi _i$ with $i=1,2,3$ for the space components of $\psi
_\mu $, while for gamma matrices $\gamma ^i$ are space components
of flat $\gamma ^a$, and similarly for the $0$ index.}
\cite{Corley} into
  its  transverse part
$\psi^T_i$, the trace $\gamma^i\psi_i$ and  the trace ${\bf k
\cdot \psi}$:
\begin{equation}
 \psi_i = \psi^T_i +\left(\frac{1}{2}\gamma_i-\frac{1}{2}{\hat k_i}
 ({\bf{\hat k} \cdot \gamma}) \right) \gamma^j \psi_j +
\left(\frac{3}{2}{\hat k_i}-\frac{1}{2} \gamma_i ({\bf{\hat k}
\cdot \gamma}) \right) { \bf{\hat k} \cdot \psi} \ , \label{split}
\end{equation}
where  ${\hat k_i}=  k_i/{\vert {\bf k }\vert}$, so that $\gamma^i
\psi^T_i={\hat k^i}\psi^T_i=0$. We will relate $\gamma^i\psi_i$
with $\psi_0$ and with $\bf{\hat k} \cdot \psi$, so that, after
use of the field equations there are two degrees of freedom
associated with the transverse part $\psi^T_i$, which correspond
to helicity $\pm3/2$; and two degrees of freedom associated with
 $\gamma^i\psi_i$ (or $\psi_0$)
which correspond to helicity $\pm1/2$.

For the helicity $\pm 3/2$ states we have to derive the equation
for $\psi^T_i$. We apply decomposition (\ref{split}) to the master
equation (\ref{master}) for $\mu=i$ and obtain\footnote{A similar
equation obtained in \cite{maz} has a different coefficient in the
term $\frac{ a'}{2a}\gamma^0$ since they have omitted the
Christoffel symbols in the covariant derivative. One can still use
their equation if one replaces the curved space gravitino vector
$\psi_\mu$ for which the equation was derived by the tangent space
vector $\psi_a= e_a^\mu \psi_\mu$. }
\begin{equation}
\left(\gamma ^a \delta _a^\mu
\partial _\mu + \frac{ a'}{2a}\gamma^0 +
 ma \right)\psi_i^T=0 \,.
\label{trans}
\end{equation}
 In the limit of
vanishing gravitino mass, the transverse part $\psi^T_i$ is
conformal with a weight $+1/2$. The transformation
$\psi^T_i=a^{-1/2}\Psi^T_i$ reduces
 the equation for the transverse part to the
free Dirac equation with a time-varying mass term $ma$. It is well
known how to treat this type of equations (e.g.\ see \cite{GK}).
The essential part  of $\Psi^T_i$ is given by the time-dependent
part of the eigenmode  of the transversal component $y_T(\eta)$,
which obeys second-order equation
\begin{equation}
y_T''+\left(k^2+\Omega_T^2-\rmi\Omega_T'\right)y_T=0 \ ,
 \label{dirac}
\end{equation}
where the effective mass is $\Omega_T=m(\eta)a(\eta)$.

The corresponding equation for gravitino with helicity 1/2 is more
complicated. We have to find $k^i\psi_i$ and $\gamma ^i \psi_i$.
  The equation for
the components $k^i\psi_i$ can be obtained from the constraint
equation (\ref{constr3}):
\begin{equation}
\rmi{\bf k \cdot \psi}=\left( -\frac{ a'}{a}\gamma_0 +\rmi \gamma
\cdot{\bf k}-ma \right)\gamma ^i \psi_i \,. \label{long}
\end{equation}
Combining all terms together, we obtain the on-shell decomposition
for the longitudinal part
\begin{equation}
  \psi _i=\psi _i^T+\left( \hat{k}_i\,\gamma \cdot \hat{\bf
k}+\frac{\rmi}{2\mathbf{k}^2}
  (3k_i-\gamma _i \mathbf{k}\cdot \gamma )(\frac{a'}{a}\gamma _0
+ma)\right)
  \gamma ^j\psi _j\,.
\label{decomppsi}
\end{equation}
Now we can derive an equation for $\gamma ^j\psi _j$. {}From the
zero component of (\ref{master})  we have
\begin{equation}
\frac{3a'}{2a}\gamma ^0\psi_{0}+\left(\frac{3}{2}m a+ \rmi {\bf k
\cdot \gamma}\right) \psi_{0}= (\gamma ^i
\psi_i)'-\frac{ma}{2}\gamma _0 \gamma ^i \psi_i \,. \label{0}
\end{equation}
This equation does not contain the time derivative of $\psi_{0}$.
Substituting $\psi_{0}$ from  (\ref{A}) into (\ref{0}),
 we get an
equation for $\gamma^i\psi_i$
\begin{equation}
 \left( \partial _\eta +\hat{B}-
\rmi{\bf k  \cdot  \gamma} \gamma _0 \hat{A} \right)
 \gamma^i \psi_i=0\,,
 \label{trace1}
\end{equation}
where
\begin{equation}
 \hat  A =-\exp \left(2{\gamma_0}\int_{-\infty }^t dt \,\mu
(\eta)\right)
 \,,
\label{hatA}
\end{equation}
and
\begin{equation}
 \hat  B =-\frac{3a'}{2a}\hat{A}-{ {m a} \over 2}\gamma _0(1
+3\hat{A}) \,.
\label{hatB}
\end{equation}
We can split the spinors $  \gamma ^i\psi$ in eigenvectors of
$\gamma _0$, $ \gamma ^i\psi _i=\theta _++\theta _-,$ and $ \theta
_\pm ={1\over 2}
  (1\mp {\rm i}  \gamma _0)\gamma ^i\psi _i$. From the Majorana
condition
it follows that  $ \theta _\pm(k) ^*=\mp {\cal C}\theta _\mp(-k),$
where ${\cal C}$ is the charge conjugation matrix. In a
representation with diagonal $\gamma _0$ the components $\theta
_\pm $ correspond to the $\gamma _0$-eigenvalues $\pm {\rm i}$.
 Acting
on (\ref{trace1}) with the hermitian conjugate operation gives us
a second-order differential equation on the $\theta _+$. We choose
for each $k$ a spinor basis $u_{1,2}(k)$
 for the two components of $\theta _+$, and two independent
solutions of the second-order differential equations
$f_{1,2}(k,\eta )$.  The general solution is given by
\begin{eqnarray}
\theta _+&=&
  \sum_{\alpha ,\beta =1}^2a^{\alpha \beta }(k)f_\alpha(k,\eta )
  u_\beta(k)\,, \nonumber\\
\theta _-&=&  \frac{A^*}{|A|^2}\frac{k\cdot \gamma }{k^2}
  \left( \partial _\eta +B\right)\theta _+
= -{\cal C}^{-1}
  \sum_{\alpha ,\beta =1}^2a^{*\alpha \beta }(-k)f^*_\alpha(-k,\eta )
  u^*_\beta(-k)\,.
\label{gensolL}
\end{eqnarray}
 The last equality
determines reality properties of the coefficients.  Here we represented
$\hat B$ as $B_1 +\gamma_0 B_2$ and defined $B = B_1 +\rmi B_2$, by analogy
with the definitions for the matrix $\hat A$. By the
substitution $f_{\alpha}(k,\eta)= E(\eta)y_L(\eta)$, with
$E=(-A^*)^{1/2} \exp\left(-
 \int^\eta d\eta\, \Re B\right),$
  equation   for the functions $f_{\alpha}(k,\eta)$ is reduced to the
final oscillator-like equation for the time-dependent mode
function $y_L(\eta)$:
\begin{equation}
y_L''+\left(k^2+\Omega_L^2-\rmi\Omega_L'\right)y_L=0 \,.
 \label{final}
\end{equation}
Here
\begin{equation}
a^{-1} \Omega_L=\mu-\frac{3}{2} H\sin 2{\int \mu dt }+ \frac{1}{2}
m \left(1 -3 \cos{2\int \mu dt }\right) \,.
 \label{eff}
\end{equation}
In the derivation of (\ref{final}) it was essential that $A$ has the
form (\ref{Astarmu}).

 Finally we give the expression for
the energy density of the longitudinal mode
\begin{equation}
a^3 T_0{}^0=- \frac{3}{4\mathbf{k}^2}\bar \psi
^i(-{\mathbf{k}})\gamma _i
 \left[\rmi \gamma \cdot \mathbf{k}\left( ma-\frac{a'}{3a}\gamma
_0\right)
 -\left( ma-\frac{a'}{a}\gamma _0\right)^2\right]
  \left( ma+\frac{a'}{a}\gamma _0\right)\gamma ^j\psi
_j(\mathbf{k})\,. \label{T00L}
\end{equation}
In the flat spacetime limit $A_1=-1$,  $\gamma ^j\psi _j \sim
\frac{\mathbf{k}}{m_0}$. {}From $ T_0{}^0$  we can define the
occupation number of gravitinos $n_k$ of energy $\omega_k$ at
given mode
 $T_0{}^0=\int d^3k \omega_k n_k$, where $n_k$ is expressed through
a bilinear combination of mode functions $y_L(\eta)$.

\section{Gravitino production}

One could expect that the gravitino production may begin already
at the stage of inflation, due to the breaking of conformal
invariance. However, there is no massive particle production in de
Sitter space (i.e. as long as one can neglect the motion of the
scalar field during inflation). Indeed, expansion in de Sitter
space is in a sense fictitious; one can always use coordinates in
which it is collapsing or even static. An internal observer living
in de Sitter space would not see any time-dependence of his
surroundings caused by particle production; he will only notice
that he is surrounded by particle excitations at the Hawking
temperature $H/2\pi$.

 Gravitino production may occur at the stage of inflation
due to the (slow) motion of the scalar field, but the most
interesting effects occur at the end of inflation, when the scalar
field $\phi$ rapidly rolls down toward the minimum of its
effective potential $V(\phi)$ and oscillates there. During this
stage the vacuum fluctuations of the gravitino field are
amplified, which corresponds to the gravitino production (in a
squeezed state).

Production of  gravitinos with helicity 3/2  is described in terms
of the mode function $y_T(\eta)$. This function  obeys the
equation (\ref{dirac}) with $\Omega_T=ma$, which is suppressed by
$M_P^{-2}$. Non-adiabaticity of the effective mass
$\Omega_T(\eta)$ results in the departure of $y_T(\eta)$ from its
positive frequency initial condition $e^{ik\eta}$, which can be
interpreted as particle production. The theory of this effect is
completely analogous to the theory of production of usual fermions
of spin 1/2 and mass $m$ \cite{GK}. Indeed, Eq. (\ref{dirac})
coincides with the basic equation which was used in \cite{GK} for
the investigation of production of Dirac fermions during
preheating.

The description of production of gravitinos with helicity 1/2 is
similar but somewhat more involved. The wave function of the
helicity 1/2 gravitino is a product of the  factor $E(\eta)$ and
the function $y_L(\eta)$. The factor $E(\eta)$ does not depend on
momenta and controls only the overall scaling of the
solution. It is the function $y_L(\eta)$ that controls particle
production which occurs because of the non-adiabatic variations of
the effective mass parameter $\Omega_L(\eta)$. The function
$y_L(\eta)$ obeys the equation (\ref{final}) with the effective
mass $\Omega_L(\eta)$, which is given by the superposition
(\ref{eff}) of all three mass scales in the problem: $\mu$, $H$
and $m$.

 In different models of the inflation, different terms of
$\Omega_L$ will have different impact on the helicity 1/2
gravitino production. The strongest effect usually comes from the
largest mass scale $\mu$, if it is varying with time. This makes
the production of gravitinos of helicity 1/2 especially important.

To fully appreciate this fact, one should note that if instead of
considering supergravity one would consider SUSY with the same
superpotential, then the goldstino $\chi$ (which is eaten by the
gravitino in supergravity) would have the mass $\partial_\Phi
\partial_\Phi W$, which coincides with $\mu$ in the limit of large
$M_P$, see Eq. (\ref{gm}). As a result,  Eq. (\ref{final})
describing creation of gravitinos with helicity 1/2 at $\phi \ll
M_P$ looks exactly as the equation describing creation of
goldstinos in SUSY. That is why production of gravitinos with
helicity 1/2 may be very efficient: in a certain sense it is not a
gravitational effect. (On the other hand, the decay rate of
gravitinos $\Gamma \sim m^3/M_P^2$ is very small because it is
suppressed by the gravitational coupling $M_P^{-2}$.)

This does not mean that one can always neglect terms proportional
to $H$ and $m$ as compared to $\mu$, and that production of
gravitinos with helicity 3/2 can always be neglected. In order to
understand the general picture, we will consider several toy
models where the effective potential at the end of inflation has
simple shape such as $V \sim \phi^n$. We will not discuss here the
problem of finding superpotentials which lead to such potentials
(and inflation) at $\phi > M_P$ \cite{book}, because we are only
interested in what happens after the end of inflation, which
occurs at $\phi \sim M_P$.

First consider the superpotential $W = \ft12m_\phi \Phi^2$. At $\phi
\ll M_P$ it leads to the simple quadratic potential $V ={
m_\phi^2\over 2} \phi^2$. The parameter $\mu$ in this case
coincides with the inflaton mass $m_\phi$. In a realistic
inflationary model one should take $m_\phi \sim 10^{13}$ GeV,
which is equal to $5\times 10^{-6} M_P$ \cite{book}. Hubble
constant during the field oscillations is given by ${m_\phi
\phi_0\over \sqrt{6} M_P}$, where $\phi_0(t)$ here is the
amplitude of the field oscillations, which decreases during the
expansion of the universe. The gravitino mass is given by $m =
{m_\phi \phi^2\over 4 M_P^2}$.

Thus, at the end of inflation in this model, which occurs at $\phi
\sim M_P$, all parameters determining the behavior of the
gravitino wave function are of the same order, $\mu \sim m_\phi
\sim H \sim m$. However, later the amplitude of $\phi$ decreases
as $\phi_0 \sim {1.5 M_P\over m_\phi t} \sim {M_P\over 4 N}$,
where $N$ is the number of oscillations of the field $\phi$ after
the end of inflation \cite{KLS}. Thus already after a single
oscillation there emerges a hierarchy of scales, $\mu \sim m_\phi
\gg H \gg m$.

Since $m_\phi = const$, after the first oscillation the parameter
$\mu$ becomes nearly constant,  the parameters $H$ and $m$ become
very small, and their contribution to the gravitino production
becomes strongly suppressed. As a result, the dominant
contribution to the gravitino production in this model occurs
within the first oscillation of the scalar field after the end of
inflation. Each of the parameters $\mu$, $H$, and $m$ at the end
of inflation changes by $O(m_\phi)$ within the time
$O(m_\phi^{-1})$. This means that (because of uncertainty
relation) gravitinos of both helicities will be produced, they
will have physical momenta $k = O(m_\phi)$, and their occupation
numbers $n_k$ will be not much smaller than $O(1)$. This leads to
the following conservative estimate of the number density of
produced gravitinos: $n_{3/2} \sim 10^{-2} m^3_\phi$.

Now let us assume for a moment that all energy of the oscillating
field $\phi$ transfers to thermal energy $\sim T^4$ within one
oscillation of the field $\phi$. This produces gas with entropy
density $s \sim T^3 \sim \left({m_\phi^2 M_P^2\over
2}\right)^{3/4}$. As a result, the ratio $n_{3/2}$ to the entropy
density becomes
\begin{equation}
{n_{3/2}\over s} \sim 10^{-2} \left({m_\phi\over M_P}\right)^{3/2}
\sim 10^{-10} \ . \label{final1}
\end{equation}
This violates the bound ${n_{3/2}\over s} < 10^{-14}$ for the
gravitino with $m \sim 10^2$ GeV by about 4 orders of magnitude.
Thus one may encounter the gravitino problem even if one neglects
their thermal production.

In this particular model one can overcome the gravitino problem if
reheating and thermalization occurs sufficiently late. Indeed,
during the post-inflationary expansion the number  density of
gravitinos decreases as $a^{-3}$. The energy density of the
oscillating massive scalar field $\rho = m_\phi^2 \phi_0^2(t)/2$
also decreases as $a^{-3}$. But the entropy produced at the moment
of reheating is proportional to $\rho^{3/4}$, so it depends on the
scale factor at the moment of reheating as $a^{-9/4}$. If
reheating occurs late enough (which is necessary anyway to avoid
thermal production of gravitinos), the ratio ${n_{3/2}\over s}
\sim 10^{-10} a^{-3/4}$ becomes small, and the gravitino problem
does not appear.

 But this simple resolution is not possible in some other models.
  As an example, consider the model with the
superpotential $W= \sqrt{\lambda}\Phi^3/3$, which at  $\phi \ll
M_P$ leads to the effective potential $\lambda\phi^4/4$. The
oscillations of the scalar field near the minimum of this
potential are described by elliptic cosine, $\phi(\eta)={\phi_0
\over a}cn(\sqrt{\lambda}\phi_0, {1 \over \sqrt{2}})$. The
frequency of oscillations is $0.8472 \sqrt{\lambda}\phi_0$ and
initial amplitude $\phi_0\simeq M_P$ \cite{KLS}.

The parameter $\mu$  for this model is given by
$\mu=\sqrt{2\lambda} \phi$. It rapidly changes in the interval
between $0$ and $\sqrt{2\lambda} \phi_0$. Initially it is of the
same order as $H$ and $m$, but then $H$ and $m$ rapidly decrease
as compared to $\mu$, and therefore the oscillations of $\mu$
remain the main source of the gravitino production. In this case
production of gravitinos with helicity 1/2 is much more efficient
than that of helicity 3/2.

The theory of production of gravitinos with helicity 1/2 in this
model is similar to the theory of production of spin 1/2 fermions
with mass $ \sqrt{2\lambda}\phi$ by the coherently oscillating
scalar field in the theory $\lambda\phi^4/4$. This theory has been
investigated in \cite{GK}. The result can be formulated as
follows. Even though the expression for $\Omega$ contains a small
factor $\sqrt{2\lambda}$, one cannot use the perturbation
expansion in ${\lambda}$. This is because the frequency of the
background field oscillations is also proportional to
$\sqrt{\lambda}$. Growth of fermionic modes (\ref{final}) occurs
in the non-perturbative regime of parametric excitation. The modes
get fully excited with occupation numbers $n_p \simeq 1/2$ within
about ten oscillations of the field $\phi$, and the width of the
parametric excitation of fermions in momentum space is about
$\sqrt\lambda\phi_0$. This leads to the following estimate for the
energy density of created gravitinos,
\begin{equation}\label{dens}
\rho_{3/2} \sim (\sqrt\lambda \phi_0)^4 \sim \lambda V(\phi_0)\ ,
\end{equation}
and the number density of gravitinos
\begin{equation}\label{numberdens}
n_{3/2} \sim   \lambda^{3/4} V^{3/4}(\phi_0)  \,.
\end{equation}

Now let us suppose that at some later moment reheating occurs and
the energy density $V(\phi_0)$ becomes transferred to the energy
density of a hot gas of relativistic particles with temperature $T
\sim V^{1/4}$. Then the total entropy of such particles will be $s
\sim T^3 \sim V^{3/4}$, so that
\begin{equation}\label{numberdens2}
{n_{3/2}\over s} \sim   \lambda^{3/4}  \sim 10^{-10}  \ .
\end{equation}
This result violates the cosmological constraints on the abundance
of gravitinos with mass $\sim 10^2$ GeV by 4 orders of magnitude.
In this model the ratio ${n_{3/2}\over s}$ does not depend on the
time of thermalization, because both $n_{3/2}$ and $V(\phi_0)^3/4$
decrease as $a^{-3}$. To avoid this problem one may, for example,
change the shape of $V(\phi)$ at small $\phi$, making it
quadratic.

The situation in models with $V(\phi) \sim \phi^n$ for $n > 4$ is
even more dangerous because the energy density of the oscillating
field $\phi$ in such models decreases faster than $a^{-4}$, so the
entropy of the particles produced during reheating is suppressed
stronger than by the factor of $a^{-3}$. The later reheating
occurs in such models, the greater will be the resulting ratio
${n_{3/2}\over s}$.

The most dangerous situation occurs in the class of inflationary
models where the effective potential $V(\phi)$ does not have a
minimum, but instead monotonically decreases and becomes flat at
$\phi \to \infty$. Such models have been studied recently by many
authors \cite{PV,FKL}, which gave them many different names:
deflation, kination, and quintessential inflation. Following
\cite{FKL}, we will call them non-oscillatory (NO) models, which
reflects an unusual non-oscillatory behavior of the scalar field
after inflation. This behavior implies that the standard mechanism
of reheating does not work in NO models. Therefore until very
recently it was assumed that in such models all particles are
produced gravitationally, due to the breaking of conformal
invariance  \cite{PV}. A typical model of such type has the
effective potential which is given by $V(\phi)\approx \lambda
\phi^4/4$ at $\phi <0$, and then it rapidly vanishes as $\phi$
becomes positive, so that $V(\phi) \to 0$ at $\phi \to \infty$.

 We will leave apart the question whether it is easy to obtain
realistic versions of NO models in supergravity.
For us it is only important that in such models the parameters
$\mu$, $H$ and $m$ change by $O(\sqrt\lambda \phi_0) = O(H)$
during the time when the field $\phi$ rolls from $\phi_0 \sim M_P$
to $0$. Just as in one of the examples considered above, this
should lead to production of gravitinos with number density which
can be estimated as $\sim 10^{-2} H^3$. This number is of the same
order as the number density of all other conformally noninvariant
particles produced by gravitational effects in the scenario of
Ref. \cite{PV}. Thus, barring the subsequent dilution of
gravitinos by some late-time entropy release, one has $n_{3/2}/s =
O(1)$, which contradicts observational data by 14 orders of
magnitude.

This problem of NO models can be resolved if one assumes that the
scalar field $\phi$ interacts with some other particles $\sigma$
with a sufficiently large coupling constant $g$. This leads to
production of particles in the context of the instant preheating
scenario \cite{inst}. This mechanism is much more efficient than
the gravitational particle production studied in \cite{PV}, and
the entropy $s$ of produced particles becomes much greater than
$O(H^3)$. This leads to a strong suppression of  $n_{3/2}/s$
\cite{FKL}.

But do we really have the gravitino problem in all of these models? In
our investigation we studied only the models with one chiral multiplet.
This is good enough to show that nonthermal gravitino production may
indeed cause a serious problem, but  much more work should be done in
order to check whether the problem actually exists in realistic models
with several different chiral and vector multiplets.

First  of all, one should write and solve a set of equations involving
several multiplets. Even in the case of one multiplet it is extremely
difficult, and the results which we obtained are very unexpected. The
situation with many multiplets is even more involved. One possibility  is
to consider the limit $M_p \to \infty$, since the most interesting
effects should still exist in this limit.

But this is not the only problem to be considered. In the toy models
studied in this section  with $W = \ft12m_\phi \Phi^2$ and $W =
\ft13\sqrt\lambda \Phi^3$, the superpotential $W$  and the gravitino
mass  vanish in the minimum of the
potential at $\phi = 0$. Then after the end of oscillations
supersymmetry is restored, superhiggs effect does not occur and
instead of massive gravitinos we have ordinary chiral
fermions.
In order to study production of gravitino with helicity 1/2 with
nonvanishing mass $m \sim 10^2$ GeV one must introduce  additional terms
in the superpotential, and make sure that these terms do not lead to a
large vacuum energy density.

Models with one chiral superfield which satisfy all of these requirements
do exist. The simplest one is the Pol\'{o}nyi model with $W = \alpha
((2-\sqrt 3)M_p + \Phi)$. One can introduce various generalizations of
this model. However, potentials in all models of this type that we were
able to construct are much more complicated than the potentials of the
toy models studied in this section. In particular, if one simply adds
small terms $\alpha +\beta\Phi
+...$ to the superpotentials $\sim \Phi^2$
or $\Phi^3$, one typically finds that $V$ becomes  negative in the
minimum of the potential, which sometimes becomes shifted to the
direction $\phi_2$, where $\Phi = (\phi_1 + \rmi\phi_2)/\sqrt 2$.  This
problem can be easily cured in realistic theories with many multiplets,
which is another reason to study such models.

 It would be most important to verify, in the context
of these models, validity of our observation that the probability
of production of gravitinos of helicity 1/2 is not suppressed by
the gravitational coupling. We have found, for example, that the
ratio ${n_{3/2}\over s}$ for the gravitinos with helicity 1/2 in
the model $\lambda\phi^4/4$ is suppressed by $\lambda^{-3/4}$.
This suppression is still rather strong  because the coupling
constant $\lambda$ is extremely small in this model, $\lambda \sim
10^{-13}$. However, in such models as the hybrid inflation
scenario  all coupling constants typically are $O(10^{-1})$
\cite{hybrid}. If production of gravitinos in such models is
suppressed only by powers of the coupling constants, one may need
to take special precautions in order to avoid producing
excessively large number of gravitinos during preheating. We will
return to this question and present a more detailed description
of the effects discussed above in a separate publication
\cite{long}.

\bigskip
\section*{Acknowledgments}
It is a pleasure to thank P. Bin\'{e}truy, A. Chamseddine, S.
Dimopoulos, M Dine, G. Dvali, P. Greene, J. Ellis, D. Lyth,   A. Riotto, L.
Susskind, S. Thomas, and I. Tkachev for useful discussions. R.K.
and A.L. are grateful to the organizers of the Les Houches  School
of Physics   for the hospitality. The work of R.K and A.L. was
supported by NSF grant PHY-9870115, L.K. was supported by NSERC
and CIAR. A.V.P. thanks the Department of Physics at Stanford for the
hospitality.

\end{document}